\documentclass[twocolumn]{revtex4}
\usepackage[dvips]{graphicx}
\usepackage{latexsym}
\usepackage{amsmath}
\usepackage{amsfonts}
\usepackage{amssymb}
\usepackage{bm}
\newcommand{\fig}[2]{\includegraphics[width=#1]{#2}}
\newcommand{\jeff}{J_{\rm eff}=1/2}
\newcommand{\etal}{\textit{et. al.}}
\begin{document}
	
\title{Correlation effects and hidden spin-orbit entangled electronic order in parent and electron-doped iridates Sr$_2$IrO$_4$}	
\author{Sen Zhou$^{1}$, Kun Jiang$^2$, Hua Chen$^3$, and Ziqiang Wang$^{2}$}
\affiliation{$^1$CAS Key Laboratory of Theoretical Physics, Institute of Theoretical Physics, Chinese Academy of Sciences, Beijing 100190, China \\
and School of Physical Sciences, University of Chinese Academy of Sciences, Beijing 100049, China}
\affiliation{$^2$Department of Physics, Boston College, Chestnut Hill, Massachusetts 02467, USA}	
\affiliation{$^3$International Center for Quantum Materials and School of Physics, Peking University, Beijing 100871, China \\
and Department of Physics, Zhejiang Normal University, Jinhua 321004, China}
\date{\today}

	
\begin{abstract}
Analogs of the high-T$_c$ cuprates have been long sought after in transition metal oxides. Due to the strong spin-orbit coupling (SOC), the $5d$ perovskite iridates Sr$_2$IrO$_4$ exhibit a low-energy electronic structure remarkably similar to the cuprates. Whether a superconducting state exists as in the cuprates requires
understanding the correlated spin-orbit entangled electronic states.
Recent experiments discovered hidden order in the parent and electron doped iridates, some with striking analogies to the cuprates, including Fermi surface pockets, Fermi arcs, and pseudogap.
Here, we study the correlation and disorder effects in a five-orbital model derived from the band theory.
We find that the experimental observations are consistent with
a $d$-wave spin-orbit density wave order that breaks the symmetry of a joint two-fold spin-orbital rotation followed by a lattice translation. There is a Berry phase and a plaquette spin flux due to spin procession as electrons hop between Ir atoms, akin to the intersite SOC in quantum spin Hall insulators. The associated staggered circulating $\jeff$ spin current can be probed by advanced techniques of spin-current detection in spintronics.
This electronic order can emerge spontaneously from the intersite Coulomb interactions between the spatially extended iridium $5d$ orbitals, turning the metallic state into an electron doped quasi-2D Dirac semimetal with important implications on the possible superconducting state suggested by recent experiments.

\end{abstract}
	
\maketitle

Sr$_2$IrO$_4$ is isostructural to the cuprate La$_2$CuO$_4$ and becomes a canted AF insulator below a N\'eel temperature $T_N\simeq230$ K \cite{kim08,kim09}.
The canting of the in-plane magnetic moments tracks the $\theta\simeq 11^\circ$ staggered IrO$_6$ octahedra rotation about the $c$ axis \cite{huang94,cava94,crawford94,cao98} due to the strong spin-orbit coupling (SOC). The AF insulating state arises from a novel interplay between SOC and electron correlation most easily understood near the atomic limit. Ir$^{4+}$ has a $5d^5$ configuration.
The $5$ electrons occupy the lower $3$-fold $t_{2g}$ orbitals separated from the higher $2$-fold $e_{g}$ orbitals by the cubic crystal field $\Delta_c$.
The strong atomic SOC, $\lambda_{\rm soc}$, splits the $t_{2g}$ orbitals into a low-lying $J_{\rm eff} = 3/2$ spin-orbit multiplet occupied by $4$ electrons and a singly occupied $J_{\rm eff} = 1/2$ doublet. Assuming $\lambda_{\rm soc}$ and $\Delta_c$ are sufficiently large compared to the relevant bandwidths when Sr$_2$IrO$_4$ crystalizes, a single $J_{\rm eff}=1/2$ band is half-filled and can be driven by a moderate local Coulomb repulsion $U$ to an AF Mott insulating state \cite{kim08,kim09,jackeli09}.
The nature of the spin-orbit entangled insulating state has been studied using the localized picture based on the $\jeff$ pseudospin anisotropic Heisenberg model \cite{jackeli09,kim12,fujiyama12,perkins14, carter13}, the three-orbital Hubbard model for the $t_{2g}$ electrons with SOC \cite{watanabe10, watanabe14, arita12, hsieh12}, and the microscopic correlated density functional theory such as the LDA+U and GGA+U \cite{kim08, jin09, liu15, pliu16}. Moreover, carrier doping the AF insulating state was proposed to potentially realize a $5d$ $t_{2g}$-electron analog of the $3d$ $e_g$-electron high-T$_c$ cuprate superconductors \cite{fawang11, kim12, watanabe10, watanabe14, meng14}.

In this work, we study the hidden order in both stoichiometric and electron-doped Sr$_2$IrO$_4$ discovered recently by angle-resolved photoemission (ARPES) and scanning tunneling microscopy (STM).
In high quality undoped Sr$_2$IrO$_4$ samples, the most recent ARPES experiment \cite{torre15} was able to resolve the broad spectra in the canted AF insulator near the high symmetry point $X=(\pi,0)$ and $(0,\pi)$ observed in earlier experiments \cite{kim08, moser14, uchida14, brouet15, nie15} and reveal a degeneracy splitting of the quasiparticle (QP) dispersion, indicative of a symmetry breaking hidden electronic order.
Electron doping the AF insulator has been achieved by La substitution (Sr$_{2-x}$La$_x$IrO$_4$) \cite{torre15, li-cpl15}, oxygen deficiency (Sr$_2$IrO$_{4-\delta}$) \cite{korneta10}, and {\em in situ} potassium surface doping \cite{kim-arc-14, kim-dpg-16, dlfeng15}.
ARPES measurements showed that the collapse of the AF insulating gap
gives rise to a paramagnetic (PM) metallic state with Fermi surface pockets for bulk electron doping at $x=0.1$ \cite{torre15} and to Fermi arcs \cite{kim-arc-14} with $d$-wave like pseudogaps around $X$ \cite{torre15,kim-dpg-16} under surface doping, in striking analogy to the high-T$_c$ cuprates. A hidden electronic order that breaks the rotation, inversion, and time-reversal symmetries has been observed in {\em hole}-doped Sr$_2$Ir$_{1-x}$Rh$_x$O$_4$ by optical second harmonic generation (SHG) \cite{zhao15} and neutron scattering measurements \cite{jeong17}. However, similar experiments have not been performed in electron-doped iridates. Moreover, since the Rh substitution of the strongly spin-orbit coupled Ir in the Ir-O plane is very different than the electron-doping by La substitution in the off-plane charge reservoir layers or surface K-doping, we shall not consider the hole-doped case further in this paper. Our focus will be the hidden electronic order in the low-energy QP properties observed by ARPES and STM in undoped and electron-doped iridates.

To this end, we study the effects of correlation, SOC, and structure distortion on the spin-orbit entangled electronic states. We show that these remarkable QP properties can be described by a $d$-wave spin-orbit density wave with a circulating staggered
$\jeff$ spin current that breaks the symmetry of two-fold spin-orbital rotation followed by lattice translation. It gaps out the band touching point at momentum $X$ and generates the electron pockets in the PM phase, and splits the degenerate band near $X$ in the canted AF insulator, in remarkable agreement with experiments. The effects of disorder are studied and shown to produce the pseudogap and Fermi arcs observed under surface doping. The electronic order induces a Berry phase associated with the staggered plaquette spin-flux as electrons hop between the Ir sites via the oxygen due to spin-precession.
We argue that the hidden order has an electronic origin and can be generated spontaneously by the intersite Coulomb interactions due to the large spatial extent of the iridium $5d$ orbitals, turning the metallic phase of the iridates into an electron doped quasi-2D Dirac semimetal. These findings provide new insights and perspectives for understanding the possible emergence of a superconducting phase \cite{dlfeng15,kim-dpg-16}.

\begin{figure}
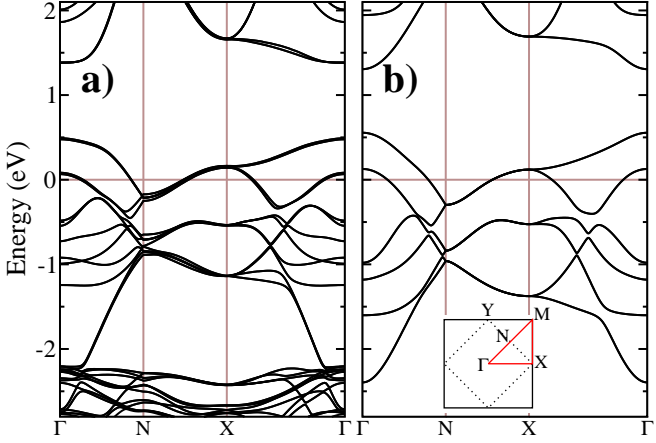

\begin{center}
\fig{3.4in}{fig1.eps}
\caption{Comparison of the band structures obtained using (a) LDA and (b) the five-orbital TB+SOC model with lattice distortion caused by staggered octahedra rotation. The doubling of bands and the small band splitting in (a) is due to the doubling of unit cell along $c-$axis. The inset in (b) shows the one-Ir Brillouin zone (solid black lines), the reduced zone (dotted black lines), and the high-symmetry points labeled by $\Gamma=(0, 0)$, $X=(\pi, 0)$, $Y=(0, \pi)$, $M=(\pi, \pi)$, and $N=(\pi/2,\pi/2)$.}
\end{center}
\end{figure}

We start with a realistic electronic structure calculation using the local density approximation (LDA) including SOC and the structural distortion \cite{martins11,QE09}.
The result is shown in Fig.~1a for $\theta=11^\circ$. We then construct a two-dimensional tight-binding model including SOC (TB+SOC) for the low-energy band structure using five localized Wannier orbitals
to be labeled by $\mu= 1 (d_{YZ}), 2 (d_{ZX}), 3(d_{XY}), 4 (d_{3Z^2-R^2}), 5 (d_{X^2-Y^2})$.
The TB+SOC Hamiltonian in the {\em local coordinates} that rotate with the octahedra is given by
\begin{align}
{\cal H}_0&=\sum_{ij,\mu\nu,\sigma} t_{ij}^{\mu\nu,\sigma}
d_{i\mu\sigma}^\dagger d_{j\nu\sigma} + \sum_{i\mu\sigma}\epsilon_\mu
d^\dagger_{i\mu\sigma} d_{i\mu\sigma} \nonumber \\
&+ \sum_{i,\mu\nu,\sigma\sigma^\prime}\lambda_{\rm soc} \left< \mu\left|\mathbf{L} \right|\nu\right> \cdot\left<\sigma\left| \mathbf{S}\right|\sigma^\prime\right> d_{i\mu\sigma}^\dagger d_{i\nu\sigma^\prime}.
\label{eq:tb}
\end{align}
Here $d_{i\mu\sigma}^\dagger$ creates an electron with spin-$\sigma$ in the $\mu$-th orbital at site $i$, and $t_{ij}^{\mu\nu,\sigma}$ is the spin-and-orbital dependent complex hopping integrals between sites $i$ and $j$ of up to fifth nearest neighbors given in the Supplemental Material.
The second term in Eq.~(\ref{eq:tb}) denotes the crystalline electric field (CEF)
$\epsilon_{1,\cdots,5}=(0,0,202,3054,3831)$meV with a separation of
$\Delta_c\equiv10Dq\approx 3.4$eV between the $t_{2g}$ and $e_g$ complexes.
The last term in Eq.~(\ref{eq:tb}) is the atomic SOC with $\lambda_\text{soc}=357$ meV; $\mathbf{S}$ and $\mathbf{L}$ are the spin and orbital angular momentum operators respectively, whose matrix elements $S^\eta_{\sigma\sigma^\prime} =\left<\sigma\left| S^\eta\right| \sigma^\prime \right>$ where $\eta=x,y,z$ in spin-space and $L^\eta_{\mu \nu} =\left< \mu\left|L^\eta\right|\nu\right>$
in the five $d$-orbital basis are given explicitly in the Supplemental Material.
The band dispersion produced by the TB+SOC Hamiltonian is shown in Fig.~1b, which captures faithfully the LDA band structure in Fig.~1a near the Fermi level. This is the first TB+SOC model of the first-principle electric structure of all five $d$-orbitals, which are necessary to describe quantitatively the lattice distortion and the atomic SOC.
Zooming in to low energies in Fig.~2a, it is clear that for the realistic bandwidths and CEF, the atomic SOC is insufficient to prevent two bands of predominantly $J_{\rm eff}=1/2$ and $3/2$ characters to cross the Fermi level and give rise to two Fermi surfaces (FS) shown in Fig.~2(a).

\begin{figure}
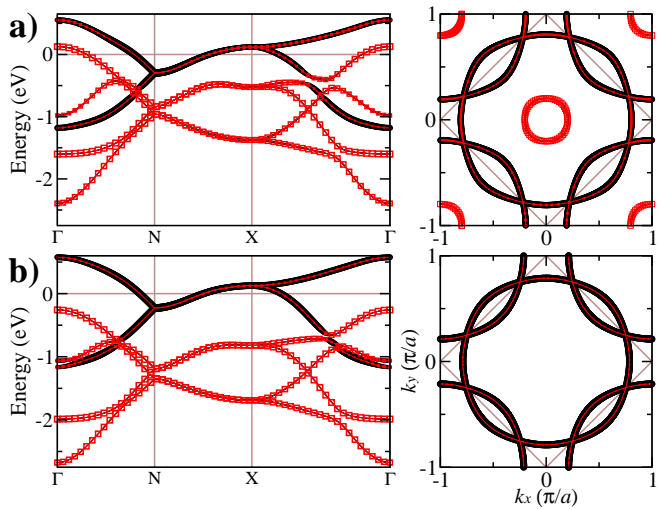

\begin{center}
\fig{3.4in}{fig2.eps} \caption{The band dispersions and the corresponding FS of (a) the noninteracting $U=0$ and (b) the nonmagnetic state in the interacting $(U,J)$=(1.2, 0.05) eV system at $x=0$. The thickness of the black lines denotes the content of the $J_\text{eff}=1/2$ doublet, while the size of the red open squares shows the content of the $J_\text{eff}=3/2$ multiplet.}
\end{center}
\end{figure}

To study the electron correlation effects, we consider the five-orbital Hubbard model ${\cal H}={\cal H}_0+{\cal H}_U$ with
\begin{align}
{\cal H}_U& = U\sum_{i,\mu}\hat{n}_{i\mu\uparrow}\hat{n}_{i\nu\downarrow}
+(U^\prime-J/2) \sum_{i,\mu < \nu}\hat{n}_{i\mu}\hat{n}_{i\nu}\nonumber\\
&-J\sum_{i,\mu\ne \nu}\mathbf{S}_{i\mu}\cdot\mathbf{S}_{i\nu} +J\sum_{i,\mu\ne\nu}d^\dagger_{i\mu\uparrow}d^\dagger_{i\mu\downarrow}
d_{i\nu\downarrow}d_{i\nu\uparrow},
\label{hu}
\end{align}
where $U$ and $U^\prime$ are the local intra- and inter-orbital Coulomb repulsions and $J$ is the Hund's rule coupling with $U = U^\prime + 2J$. In the presence of SOC, the Hartree and exchange self energies induced by ${\cal H}_U$ in the self-consistent Hartree-Fock theory depend on the full spin-orbital dependent density matrix $n^{\mu\nu}_{i\sigma\sigma'}= \langle d^\dagger_{i\mu\sigma} d_{i\nu\sigma'} \rangle$.
Local physical quantities in the ground state can be expressed in terms of $n^{\mu\nu}_{\sigma\sigma'}$.
The orbital occupation $n_\mu=\sum_{\sigma}n^{\mu\mu}_{\sigma\sigma}$,
the spin density $S^\eta=\sum_{\mu, \sigma\sigma'} S^\eta_{\sigma\sigma'} n^{\mu\mu}_{\sigma\sigma'}$,
the orbital angular momentum $L^\eta=\sum_{\mu\ne\nu,\sigma} n^{\mu\nu}_{\sigma\sigma} L_{\mu\nu}^\eta$,
and the SOC $\Lambda^\eta= \sum_{\mu\nu, \sigma\sigma^\prime} n_{\sigma\sigma^\prime}^{\mu\nu} L_{\mu\nu}^\eta S_{\sigma\sigma^\prime}^{\eta}$ can be determined from the Hartree and exchange self energies.

{\em Interacting electronic structure.} In the absence of symmetry breaking ($\langle{\bf S}\rangle =\langle{\bf L}\rangle=0$), the only corrections to the electronic structure are the changes in the CEF and the renormalization of the atomic SOC.
However, due to the cubic crystal field,
$\Lambda^\eta$ are different along different directions. As a result, the correlation induced SOC renormalization is both directional and orbital dependent, {\it i.e.} the $\lambda_{\rm so}$ in Eq.~(\ref{eq:tb}) is replaced by $\lambda_{\mu\nu}^\eta=\lambda_{\rm soc}+\Delta\lambda_{\mu\nu}^\eta$ where $\Delta\lambda\propto\lambda_{\rm so}(U^\prime -J)N_F$, where $N_F$ is the Fermi level density of states.
In Fig.~2b, the interacting electronic structure is shown at $(U,J)$=(1.2, 0.05) eV and $x=0$.
We find that the most important correlation effect on the electronic structure is the renormalization of $\lambda_{\rm soc}$,
leading to a significantly enhanced effective SOC of 665 meV for the $t_{2g}$ complex much larger than the bare atomic value. As a consequence, the $J_{\rm eff}=3/2$ band in the LDA band structure in Fig.~2a is pushed below the Fermi level together with the hole FS pockets around $\Gamma$ and $M$. As shown in Fig.2b, this gives rise to the single band crossing the Fermi level that is of dominant $J_{\rm eff}=1/2$ character and folded by $(\pi,\pi)$ due to the lattice distortion. This correlation induced band polarization
through enhancement of the SOC by the Hubbard interaction enables the $J_{\rm eff}=1/2$ picture.

\begin{figure}
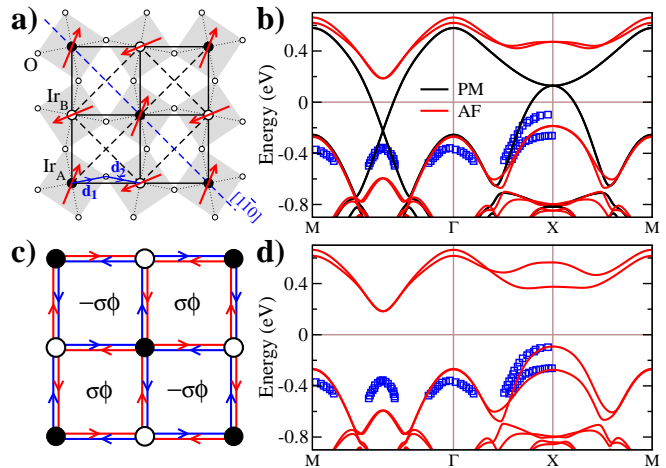

\begin{center}
\fig{3.4in}{fig3.eps}
\caption{(a) Schematics of the in-plane canted AF moments on the Ir square lattice with a two-dimensional rendering of the staggered IrO$_6$ octahedral rotation. (b) QP band dispersion in the canted AF state (red solid lines) and the nonmagnetic state (black solid lines) at $(U,J)$=(1.2, 0.05) eV and $x=0$. (c) Schematics of the $d$-wave $\jeff$ pseudospin current order and associated staggered pseudospin flux $\Phi_\sigma=\pm\sigma\phi$. (d) QP band dispersion in the canted AF state with coexisting $d$-wave $\jeff$ pseudospin current corresponding to $\Delta_d=30$meV. Open blue squares are data from ARPES experiments \cite{torre15}.}
\end{center}
\end{figure}

{\em Canted AF insulator at $x=0$.}
The fully self-consistent Hartree-Fock ground state at nonzero values of $U$ and $J$ is indeed a canted AF insulator as depicted in Fig.~3a with fully gapped QP dispersions shown in red-solid lines in Fig.~3b.
The magnetization and the canting angle depend on the interaction parameters. For $(U,J)=(1.2,0.05)$eV, the ordered magnetic moment is $\vert \langle {\bf L}\rangle +2\langle{\bf S}\rangle \vert\simeq 0.67\mu_B$ with a canting angle of about $22^\circ$, which are larger than the experimental values. Note that weak-coupling approaches tend to overestimate magnetism; the results should be regarded as qualitative rather than quantitative \cite{angle} and our findings will not rely on these microscopic details.
Comparing to the observed QP dispersion below the Fermi level in Fig.~3b reveals an important difference near the high symmetry point $X$: ARPES detects two split bands separated by about $200$ meV as shown by the superimposed blue squares. A symmetry analysis is necessary to understand this difference. In the PM state (black lines), there exists a band touching point (BTP) located at $X$ about $130$meV above the Fermi level. This is a four-fold degenerate van Hove crossing point of the $\jeff$ doublet band with its $(\pi,\pi)$-folded counter part by the structural distortion. The SOC does not affect the band degeneracy at $X$. Although the AF order breaks the time-reversal (${\cal T}$) symmetry and splits the Kramers pair, the QP band below the AF gap in Fig.~3b still maintains a two-fold degeneracy. The two split bands observed by ARPES near $X$ are thus consistent with a $\pm100$ meV band degeneracy lifting due to an additional symmetry breaking associated with a hidden order in the canted AF state.

{\em Symmetry analysis: hidden order in AF phase.} The symmetry implications on the electronic structure in Sr$_2$IrO$_4$ are subtle because of the strong SOC that renders the point group incomplete.
In the absence of SOC, the doubly-degenerate band in a two-sublattice collinear AF state
is related to a symmetry operation that flips the spin
followed by an $A\leftrightarrow B$ sublattice translation $\tau_{AB}$ \cite{kivelson}.
In the iridates, spin rotations are coupled to orbital spatial rotations due to the strong SOC.
As a consequence, flipping the spin by the $180^\circ$ spin-rotation $R_s^\prime(\pi)$ around the $[1\bar{1}0]$ axis (direction of the canted FM moment)
in Fig.~3a must be accompanied by a spatial $C_2^\prime$ rotation around the same axis, leading to the joint two-fold rotation ${\cal J}_2^\prime\equiv R_s^\prime(\pi)\otimes C_2^\prime$.
The corresponding operation in the spin-orbit entangled states must therefore be extended to ${\cal R}_2^\prime\equiv {\cal J}_2^\prime\otimes \tau_{AB}=R_s^\prime(\pi)\otimes C_2^\prime\otimes\tau_{AB}$.
${\cal R}_2^\prime$ is indeed a symmetry {\em even in the presence of spin canting and the structural distortion}. Thus, a hidden order that breaks ${\cal R}_2^\prime$ would lift the two-fold band degeneracy. To determine the specific form of the hidden order, one can exhaust all possible interactions involving the low energy $t_{2g}$ orbitals that do not break lattice translation, identify those that break ${\cal R}_2^\prime$, and examine their momentum space anisotropy according to experiments \cite{jiang16}.
Since the low energy physics here is dominated by the $\jeff$ quantum states, the outcome can be suitably understood in the local pseudospin basis discussed in the Supplemental Material, $\vert J={1/2},J_z=\pm{1/2}\rangle=\gamma_{\pm}^\dagger \vert 0\rangle$ where
$
\gamma_\sigma= {1\over\sqrt{3}}\left( i\sigma d_{YZ,\bar\sigma}+d_{ZX,\bar\sigma}+id_{XY,\sigma}\right)
$
creates the $\jeff$ doublet in QP excitations.
Under the joint two-fold spin-orbital rotation ${\cal J}_2^{\prime\dagger} \sum_\sigma\sigma \gamma_\sigma^\dagger \gamma_\sigma {\cal J}_2^\prime= -\sum_\sigma \sigma\gamma_{\sigma}^\dagger \gamma_{\sigma}$, {\it i.e.} the pseudospin is flipped, as does its current.
Considering the $C_2^\prime$ rotation and the sublattice translation $\tau_{AB}$, we arrive at the desired degeneracy lifting interaction 
\begin{equation}
{\cal H}_{\Delta} = i \Delta_d\sum_{i\in A,\sigma}\sum_{j=i+\delta} (-1)^{i_y+j_y} \sigma \gamma^\dagger_{i,\sigma} \gamma_{j,\sigma} + h.c.
\label{spincurrent}
\end{equation}
where $\delta=\pm \hat x, \pm\hat y$ and $(-1)^{i_y+j_y}$ is the standard nearest neighbor (nn) $d$-wave form factor.
${\cal H}_\Delta$ maintains ${\cal T}$, but breaks ${\cal R}_2^\prime$ since ${\cal R}_2^{\prime\dagger} {\cal H}_\Delta {\cal R}_2^\prime=-{\cal H}_\Delta$.
Eq.~(\ref{spincurrent}) describes staggered ($d$-wave) circulating $\jeff$ {spin currents} around each plaquette. Physically, the QPs traverse around a plaquette acquire a Berry phase from the enclosed spin flux for each pseudospin component in opposite directions as shown in Fig.~3c.
A nonzero expectation value of ${\cal H}_\Delta$ in the ground state thus gives rise to the $\jeff$ $d$-wave pseudospin current order (d-PSCO). Note that Eq.~(\ref{spincurrent}) can also be interpreted as a $\jeff$ $d$-wave spin-orbit density wave (d-SODW) that can exhibit long-range order without the ${\cal T}$-breaking magnetic or charge current order. We will use d-PSCO and d-SODW interchangeably. Moreover, ${\cal H}_\Delta$ also has the form of a $d$-wave spin-orbit coupling of the $\jeff$ QP and the $d$-wave form factor is crucial for breaking ${\cal R}_2^\prime$.
Indeed, projecting the tight-binding ${\cal H}_0$ into the $\jeff$ basis in the local coordinates generates an extended $s$-wave SOC due to the structural distortion \cite{fawang11}, which is invariant under ${\cal R}_2^\prime$ and already present in the band theory.

Including ${\cal H}_\Delta$ in the total Hamiltonian ${\cal H}={\cal H}_0+{\cal H}_U+{\cal H}_\Delta$, the canted AF phase at $x=0$ indeed coexists with d-PSCO for remarkably small $\Delta_d$.
The calculated QP dispersion is shown in Fig.~3d for $\Delta_d=30$ meV, which produces a $\pm100$ meV degeneracy splitting at $X$ in remarkable agreement with experiments \cite{torre15}. The nn $\jeff$ QP correlator $\chi_{ij}^\sigma=\langle \gamma_{i\sigma}^\dagger\gamma_{j\sigma}\rangle= \chi_{ij}^\prime+i\sigma \chi_{ij}^{\prime\prime}$ and the staggered pseudospin flux $\Phi_\sigma=\pm\sigma\phi$, $\phi=\sum_\square \tan^{-1}(\chi_{ij}^{\prime\prime}/\chi_{ij}^\prime)\simeq 0.055\pi$. 
The QP number current on a link is $J_{ij}^\sigma\propto{\rm Im} \chi_{ij}^\sigma$, such that the pseudospin current $J_{ij}^{\rm ps}=\sum_\sigma \sigma J_{ij}^\sigma\ne0$, whereas the charge current $J_{ij}=\sum_\sigma J_{ij}^\sigma=0$, giving rise to the novel ${\cal T}$-symmetric d-SODW state with $d$-wave pseudospin current. The $d$-wave form factor ($\cos k_x-\cos k_y$) ensures splitting is largest at $X$, and vanishes along the $\Gamma-N-M$ path in the BZ in agreement with the superimposed ARPES data. The mismatch in the band positions near the $N$ point is because the band crossing in the PM phase (black lines in Fig.3b) is too far below the Fermi level in the weak-coupling theory. The correlation induced band narrowing in the strong coupling treatment of $U$, such as in the Gutzwiller approximation or the slave boson approaches, would significantly reduce this quantitative discrepancy.

A closer look at the calculated band dispersion near $X$ in Fig.~3d reveals a weak asymmetry between the directions $\Gamma\to X$ and $M\to X$ (equivalent to $\Gamma\to Y$ in 2D) related by $90^\circ$ rotations. In the absence of structure distortion, although the magnetic ordered state and the $\jeff$ d-PSCO break the four-fold rotation symmetry ($C_4$ plus $90^\circ$ spin rotation due to SOC), nematic asymmetry in the electronic dispersion should be absent since
${\cal J}_2^\prime\otimes {\cal T}$ remains a good symmetry
that effectively interchanges $k_x\leftrightarrow k_y$.
Thus, the weak asymmetry in the QP dispersions near $X$ is due to the structure distortion that breaks ${\cal J}_2^\prime\otimes {\cal T}$, causing a mixing the $s$-wave SOC with the $d$-wave pseudospin current on the bonds. Direct measurement of the QP dispersion along $\Gamma\to Y$ by ARPES would be very desirable for comparison in Fig.~3d in order to verify the weak nematicity induced by the proposed d-PSCO.

The ability of the d-PSCO to split the band degeneracy around X offers insights into the effects of disorder. A significant source of disorder that affects the sample quality even in the undoped Sr$_2$IrO$_4$ is the disorder in the structural distortion, in particular the spatially inhomogeneous variations in the staggered IrO$_6$ octahedral rotation and thus those in the $s$-wave SOC for the $\jeff$ QP \cite{fawang11} about the average value. This necessarily generates a local distribution of the d-PSCO with ``smeared out'' band splittings that contribute to the broad spectrum near $X$ observed in all ARPES measurement \cite{kim08, moser14, uchida14, brouet15, nie15, torre15}. The band splitting was resolved recently in presumably better quality samples \cite{torre15}.

\begin{figure*}
\begin{center}
\fig{7in}{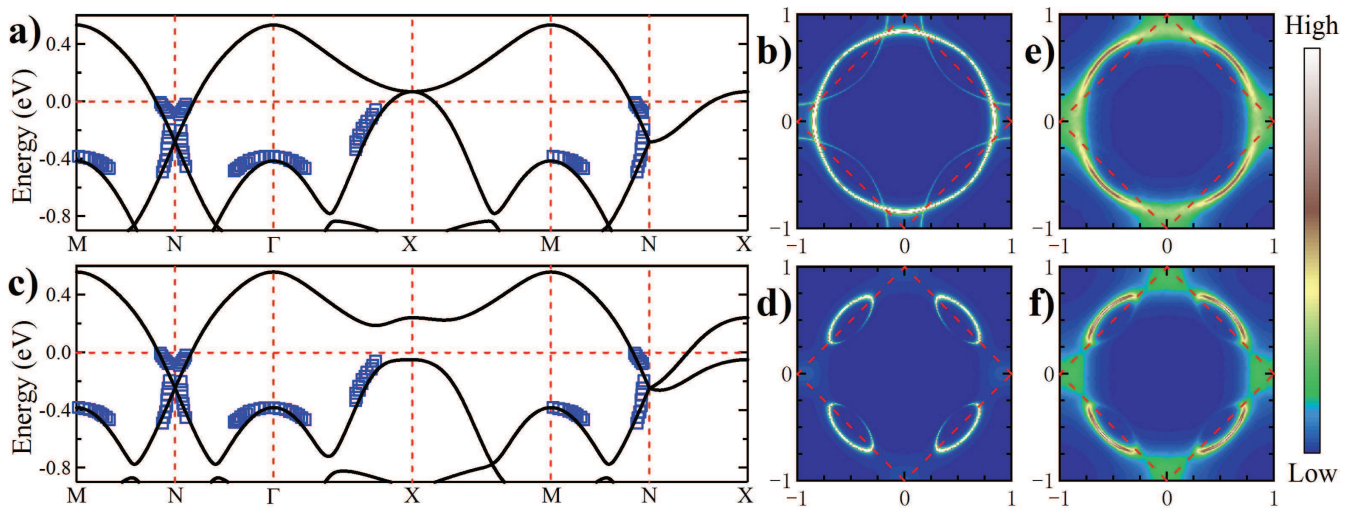}
\caption{The PM phase at $x=0.1$ and $(U, J)$=(1.4, 0) eV.(a-b) Band dispersion in the absence $H_\Delta$, i.e without d-PSCO (black lines) and the spectral intensity of the corresponding FS. (c-d) Band dispersion in the presence of d-PSCO for $\Delta_d=30$meV showing the gapping of the BTP at $X$ (red lines) and the spectral intensity of the corresponding FS. (e) FS spectrum in the presence of disorder with $\Delta_d=0$ and $\varepsilon=60$meV, showing the Fermi arcs. (f) FS spectrum in the disordered d-PSCO with $\Delta_d=30$meV and $\varepsilon=60$meV. ARPES data \cite{torre15} are superimposed as open blue squares.}
\end{center}
\end{figure*}

{\em Electron doped PM state.}
It is remarkable that such a novel spin-orbit entangled order in the canted AF phase can also account for the Fermi pocket, Fermi arc, and pseudogap phenomena in the PM phase following the collapse of the AF insulating gap in electron doped Sr$_2$IrO$_4$.
Indeed, we could have started the discussion with the electron doped case, as shown below, and arrive at the same conclusion for the $d$-wave SODW order. More detailed symmetry analysis is given in the supplemental section D.

The calculated QP band dispersions in the PM state without ${\cal H}_\Delta$ are shown in Fig.~4a for $(U, J)$=(1.4, 0) eV at $x=0.1$.
Despite the absence of AF order, the bands are still folded by $(\pi,\pi)$ due to the staggered IrO$_6$ octahedra rotation.
Moreover, {\em electron} doping has moved the Fermi energy ($E_F$) upward to within 70meV of the BTP at $X$ in Fig.~4a.
Comparing to the superimposed dispersion measured by ARPES \cite{torre15}, the electron FS pocket around $(\pi/2,\pi/2)$ in Fig.~4b is indeed observed with the QP dispersion extending from $E_F$ down to and beyond the ``Dirac crossing'', consistent with the calculated QP dispersion in Fig.~4a.
However, the hole FS pocket around $X$ in Fig.~4b was not observed by ARPES; the measured QP peak near $X$ follows the band dispersion below $E_F$ but loses its intensity before reaching the Fermi level, leading to the emergence of a $\sim30$ meV gap at the $X$ point \cite{torre15}.
It is natural to suspect that
short-range AF order or fluctuations \cite{liu16,gretarsson16} may be responsible for the (pseudo)gap behavior \cite{jxli15}. However, the latter would produce a significant energy gap in the QP dispersion around $(\pi/2,\pi/2)$ as well which was not detected by ARPES \cite{torre15}.

We propose that closing the AF gap by electron doping reveals the d-PSCO already present in the canted AF phase. A gap opening at $X$ requires symmetry breaking and lifting of the degeneracy at the BTP.
The ${\cal T}$ symmetry in the PM phase protects the two-fold Kramers degeneracy. 
However, the proximity of $E_F$ to the van Hove BTP increases the propensity toward d-PSCO that spontaneously breaks the ${\cal R}_2^\prime$ symmetry and lifts the remaining two-fold degeneracy by gapping out the BTP.
Fig.~4c shows the calculated QP dispersion in the presence of ${\cal H}_\Delta$ in Eq.~(\ref{spincurrent}) with an identical magnitude $\Delta_d=30$meV used in the AF phase at $x=0$. The induced staggered pseudospin flux is $\pm\sigma\phi$ with $\phi= 0.589\pi$. The d-PSCO splits the BTP by about $200$meV and produces a $30$meV gap in the QP dispersion at $X$ shown in Fig.~4c, leaving behind only the electron pocket around $(\pi/2,\pi/2)$ in Fig.~4d occupied by the doped carriers,
in very good agreement with the superimposed ARPES data in electron-doped Sr$_{2-x}$La$_x$IrO$_4$ at $x=0.1$ \cite{torre15}.
As in the undoped case, it would be very desirable to have the measured dispersions available along $\Gamma\to Y$ for comparison to the predicted weak nematicity and along $X\to N$ for the existence of the Dirac point.
Fig.~4d shows that the QP spectral weight is much larger on the outer half than on the inner half of the FS pockets, consistent with the former being the main $\jeff$ QP band while the latter the folded band (Fig.~4c) by the structural distortion and d-SODW order.
Significant photon energy dependent spectral weight anisotropy on the electron pocket has been observed by ARPES.
The current theory, however, cannot explain the surprising result that, at certain photon energies, the folded portion has higher intensity which may be caused by the matrix element effects \cite{torre15}.

{\em Disorder effects, Fermi arc and pseudogap behavior.} We have explained that the native disorder in the structural distortion of the undoped iridates leads to local modulations of the d-PSCO and the broadening of the spectral function near $X$ in the canted AF state. Career doping
usually introduces additional sources of disorder.
In the high-T$_c$ cuprates, doping induced disorder contributes significantly to the electronic inhomogeneity \cite{pan01,zqwang02,mcelroy05,sen07,hoffman12}. Moreover,
when a system sits close to a long-range electronic order, disorder can pin the low-energy quantum fluctuations to form a spatially inhomogeneous state with glassy or short-range order that inherits certain spectroscopic properties of the ordered state. The $d$-wave valence bond glass is such an example proposed for the pseudogap phase with Fermi arcs in underdoped cuprates \cite{ren09}.
To model the disorder effects, we rewrite Eq.~(\ref{spincurrent}) in real space,
\begin{equation}
{\cal H}_{\Delta}^{\rm dis} = i \sum_{i\in A,\sigma}\sum_{j=i+\delta} \Delta_{ij} \sigma \gamma^\dagger_{i,\sigma} \gamma_{j,\sigma} + h.c.
\label{disorderspincurrent}
\end{equation}
where the bond coupling $\Delta_{ij}=(-1)^{i_y+j_y} \Delta_d +\delta \Delta_{ij}$ contains an average $d$-wave contribution $\Delta_d$ and a random $\delta \Delta_{ij}$ taken from a Gaussian distribution of zero mean and standard deviation $\varepsilon$. Note that the disordered $\delta \Delta_{ij}$ necessarily involve spatially fluctuating $\jeff$ $d$-wave and $s$-wave SODW or spin currents.
We first set $\Delta_d=0$, such that the d-PSCO vanishes on average $\langle {\cal H}_\Delta\rangle=0$, i.e. without long-range order, but its moment $\langle {\cal H}_\Delta^2\rangle\ne0$, giving rise to short-range ordered SODW or equivalently a valence bond glass of $\jeff$ spin-current. The FS obtained with quenched disorder average \cite{disorder} is plotted in Fig.~4e at $x=0.1$ for $\varepsilon=60$meV, which shows the destruction of the FS sections around $X$ as in the clean case with long-range d-PSCO.
More remarkable is the emergence of the full fledged Fermi arcs as the folded part of the FS is destroyed by the scattering due to spatially fluctuating pseudospin current in good agreement with the observed Fermi arcs and $d$-wave like pseudogaps by ARPES and STM in heavily surface K-doped iridates \cite{kim-arc-14,dlfeng15,kim-dpg-16}.
Fig.~4f is a spectral intensity plot of the FS when the disordered pseudospin currents fluctuate spatially around a nonzero mean of the d-PSCO with $\Delta_d=\pm30$ meV to account for averaging over two domains. Since the static order is comparable to the disorder strength, although the spectral weight on the inner halves of the electron pockets is suppressed, the bending over of the Fermi arcs remains visible.

{\em Discussions.} We have shown that the highly unconventional QP properties observed in both the parent and electron doped square lattice iridates can be described by the same $d$-wave spin current or SODW order proposed in Eq.~(\ref{spincurrent}). The basic mechanism is that the spin-orbit entangled electronic order breaks the ``hidden'' $R_2^\prime$ symmetry, but it is equally surprising and reassuring that the manifestations of the corresponding degeneracy lifting of the quantum states at the high symmetry point $X$ can account for the ARPES and STM observations in both the electron doped PM pseudogap phase as well as the undoped canted AF insulator.
Other symmetry-breaking interactions capable of lifting the degeneracy at $X$, such as those discussed in the supplemental section D, do not have this property. For example, the ${\cal T}$-breaking $d$-wave circulating current or the staggered flux order gaps out the BTP at $X$ in the PM phase, but cannot produce the band splitting in the AF insulator, whereas the $d$-wave spin nematic order produces the band splitting in the canted AF insulator, but cannot remove the hole Fermi surface pocket around $X$ in the electron doped PM phase. The $d$-wave bond nematic order does not split the degeneracy in the AF insulator and produces strongly nematic QP band dispersion near $X$ in the PM metallic state, incompatible with experimental findings. It is important to note that the proposed d-PSCO in Eq.~(3) is time-reversal invariant. Thus it doesn't describe the time-reversal breaking hidden order in {\em hole}-doped Sr$_2$Ir$_{1-x}$Rh$_x$O$_4$ observed by SHG \cite{zhao15} and neutron scattering measurements \cite{jeong17}. Given the different nature of the chemical doping and the extension of the magnetically ordered phase, it is very desirable for these measurements to be carried out for the electron-doped iridates.

The most direct manifestation of the d-PSCO order is the Fermi pocket and the Fermi arc/pseudogap behavior in the electron doped PM phase. However, since it breaks different symmetries than the canted AF order, two separate phase transitions are expected in the undoped and lightly electron doped AF phase, which should be observable with improved sample quality. Still, direct experimental detections of the d-PSCO would be most convincing. The circulating spin current can in principle be probed experimentally by the recent advances in spin current detection in spintronics using optical SHG \cite{werake} and x-ray magnetic circular dichroism \cite{qiu}. However, the $d$-wave or the staggered nature of the ordered current makes the detection of a ``net'' spin current or spin flux by these techniques challenging. Hence, utilizing the response of the d-PSCO to the local environment near nonmagnetic or magnetic impurities \cite{gcao15,longzhang16} may be more suitable for its detection by local probes such as NMR and STM, in addition to the above mentioned methods. It may be possible to detect the $d$-wave form factor from the spatial patterns of the local density of states accessible by STM. A distribution of net spin flux/current may also emerge and be picked up by spintronic techniques near nonmagnetic impurities, or even net charge flux/current near magnetic impurities that can be detected by local magnetometry such as an atomic force magnetometer.

The undoped Sr$_2$IrO$_4$ does exhibit a lowering of crystal symmetry at high-temperatures observed by neutron and resonant X-ray scattering
\cite{ye13,dhital13,boseggia13} and shown by optical SHG as due to the staggered tetragonal distortion of the IrO$_6$ octahedra \cite{tetragonaldistortion}.
However, the important $d$-wave factor in $H_\Delta$ requires additional $C_4$ symmetry breaking. To illustrate this point, it is instructive to consider the spin precession due to SOC when the QP hops between the nn Ir atoms via the oxygen, as indicated by the vectors $\vec d_1$ and $\vec d_2$ in Fig.~3a. This intersite SOC \cite{kanemele} is given by $i\lambda_{12}(\vec d_1\times \vec d_2)\cdot\vec\tau_{\sigma\sigma^\prime}\gamma_{1\sigma}^\dagger
\gamma_{2\sigma^\prime}$, which leads to Eq.~(\ref{spincurrent}) if $C_4$ symmetry is broken. Since in the undoped canted AF state, the SHG signals even break the $C_2$ symmetry \cite{zhao15}, the two-dimensional d-PSCO proposed here is allowed although it does not break inversion or $C_2$ within a single layer; nor does it break ${\cal T}$ already broken by magnetic order. Further studies on the $c$-axis stacking of the d-PSCO and the magnetic order \cite{norman} are necessary in order to compare directly to the nonlinear optics and the interpretation in terms of intra-cell loop currents \cite{zhao15}.

It is likely that the main driving force behind the d-PSCO has an electronic origin. The nature of the nonlocal spin current suggests that it may emerge from intersite electronic interactions. Since the $5d$ orbitals have a large spatial extent, the nn interatomic Coulomb interaction $V$ can be important. Indeed, Eq.~(\ref{spincurrent}) can be obtained by decoupling $V$ between the $\jeff$ QPs as in the study of topological Mott insulators \cite{raghu}. Our preliminary calculations using a single-band $t$-$U$-$V$ model for the $\jeff$ QPs, with band parameters extracted from the present theory, indeed show spontaneous generation of the d-PSCO above a critical $V$ both in the PM phase and in the AF phase coexisting and competing with the AF order. The nonlocal charge fluctuations of spin-orbit coupled QPs governed by the interatomic $V$, unfavorable for $d$-wave pairing via spin fluctuations or the superexchange interaction, present a crucial difference between the iridates and the cuprates.

The d-PSCO offers a new perspective on the electronic structure of the iridates since it splits the QP band degeneracy from $X$ to $N$ except for the Dirac point at $(\pi/2,\pi/2)$ (see Fig.~4c) protected by the nonsymmorphic space group symmetries \cite{kane}. Thus, the metallic state of the iridates behaves as an electron doped quasi-2D Dirac semimetal,
which may play an essential role for studying electronic pairing and the possible emergence of superconductivity in Sr$_{2-x}$La$_x$IrO$_4$ \cite{dlfeng15,kim-dpg-16}.

We thank Stephen Wilson, Xi Dai, and Jiadong Zang for useful discussions. This work is supported by the U.S. Department of Energy, Basic Energy Sciences Grant No. DE-FG02-99ER45747 (Z.W.) and the Key Research Program of Frontier Sciences, CAS, Grant No. QYZDB-SSW-SYS012 (S.Z.). Numerical calculations were performed on HPC Cluster of ITP-CAS. Z.W. thanks the hospitality of Aspen Center for Physics and the support of ACP NSF grant PHY-1066293.

\newpage
\renewcommand{\theequation}{S\arabic{equation}}
\renewcommand{\thefigure}{S\arabic{figure}}
\renewcommand{\thetable}{S\arabic{table}}
\setcounter{equation}{0}
\setcounter{figure}{0}

\section*{ Supplementary Material }
\subsection{Orbital angular momentum and the Spin-orbit coupling}

In terms of the spherical harmonics $|\ell, m\rangle \equiv Y^m_\ell$, the five atomic $d$-orbitals can be expressed \cite{wiki} as $\psi^d = V \psi^\ell$, with
\begin{align}
\psi^d &=\left( d_{YZ}, d_{ZX}, d_{XY}, d_{3Z^2-R^2}, d_{X^2-Y^2} \right)^{\rm T}, \nonumber \\
\psi^\ell &=\left( |2,-2\rangle, |2,-1\rangle, |2,0\rangle, |2,1\rangle, |2,2\rangle \right)^{\rm T}, \text{ and} \nonumber \\
V&= \left( \begin{array}{ccccc}
0 & \frac{i}{\sqrt{2}} & 0 & \frac{i}{\sqrt{2}} & 0 \\
0 & \frac{1}{\sqrt{2}} & 0 & -\frac{1}{\sqrt{2}} & 0 \\
\frac{i}{\sqrt{2}} & 0 & 0 & 0 & -\frac{i}{\sqrt{2}} \\
0 & 0 & 1 & 0 & 0 \\
\frac{1}{\sqrt{2}} & 0 & 0 & 0 & \frac{1}{\sqrt{2}}
\end{array} \right). \nonumber
\end{align}
Using the relation $L_\pm =L^x\pm iL^y$, and the angular momentum algebra $L_\pm |\ell, m\rangle=$ $\sqrt{\ell(\ell+1)-m(m\pm 1)}$ $|\ell,m\pm 1\rangle$, $L^z |\ell, m\rangle$ $=m|\ell, m\rangle$, it is straightforward to obtain the explicit matrix form of the orbital angular moment \textbf{L} in the $\psi^d$ basis,
\begin{equation}
L^x=\left( \begin{array}{ccccc}
0 & 0 & 0 & -i\sqrt{3} & -i \\
0 & 0 & i & 0 & 0 \\
0 & -i & 0 & 0 & 0 \\
i\sqrt{3} & 0 & 0 & 0 & 0 \\
i & 0 & 0 & 0 & 0
\end{array} \right), \nonumber
\end{equation}
\begin{equation}
L^y = \left( \begin{array}{ccccc}
0 & 0 & -i & 0 & 0 \\
0 & 0 & 0 & i\sqrt{3} & -i \\
i & 0 & 0 & 0 & 0 \\
0 & -i\sqrt{3} & 0 & 0 & 0 \\
0 & i & 0 & 0 & 0
\end{array} \right), \nonumber
\end{equation}
\begin{equation}
\text{   and   } L^z= \left( \begin{array}{ccccc}
0 & i & 0 & 0 & 0 \\
-i & 0 & 0 & 0 & 0 \\
0 & 0 & 0 & 0 & i2 \\
0 & 0 & 0 & 0 & 0 \\
0 & 0 & -i2 & 0 & 0
\end{array} \right).
\end{equation}

In order to derive the spin-orbit coupling (SOC), the $d$-orbital basis needs to be enlarged to include the spin degrees of freedom
\begin{equation}
\psi^d=(d_{YZ,\downarrow}, d_{ZX,\downarrow}, d_{XY,\downarrow}, d_{3Z^2-R^2,\downarrow}, d_{X^2-Y^2,\downarrow}, \downarrow \rightarrow \uparrow)^\text{T}. \nonumber
\end{equation}
The matrix form of the atomic SOC in the spin-orbital basis is thus given by
\begin{equation}
\mathcal{H}_\text{soc}=\lambda_\text{soc} \textbf{L}\cdot\textbf{S} =\frac{1}{2}\lambda_\text{soc} \left( \begin{array}{cc} -L^z & L_+ \\ L_- & L^z \end{array} \right).
\label{SOC:d}
\end{equation}
It is straightforward to show that the SOC interaction $\mathcal{H}_\text{soc}$ is invariant under joint spin-orbital rotations, \textit{i.e.}, $\mathcal{J}^\dagger (\theta) \mathcal{H}_\text{soc} \mathcal{J} (\theta)=\mathcal{H}_\text{soc}$,
where $\mathcal{J}(\theta) = \mathcal{R}_L(\theta) \otimes \mathcal{R}_S(\theta)= e^{iL_z\theta} \otimes e^{iS_z\theta}$ rotates simultaneously the orbital and the spin by an angle $\theta$ around the $c$-axis.

\subsection{LDA band structure and the TB+SOC model}

\begin{figure}
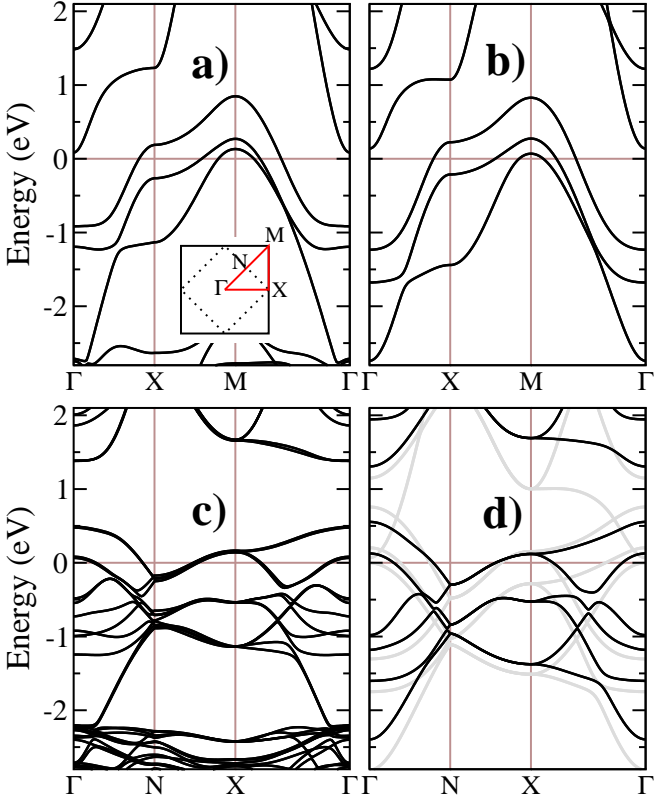

\begin{center}
\fig{3.4in}{figS1.eps}
\caption{Comparison between the LDA band structures and the TB+SOC model. (a) LDA without structural distortion.
(b) TB+SOC model without structural distortion.
(c) LDA with structural distortion.
(d) TB+SOC model with structure distortion (black lines).
The grey lines in (d) show the band structure in (b) folded into the reduced BZ.}
\label{figS1}
\end{center}
\end{figure}

We first consider an idealized, i.e. {\it undistorted} Sr$_2$IrO$_4$ where the staggered rotation of the IrO$_6$ octahedra is neglected.
The resulting crystal structure has a space-group of $I4/mmm$ symmetry and the corresponding unit cell contains only one Ir atom.
The lattice parameters are from Ref. \cite{Srandall57}: $a=b=$3.89 \AA\ and $c=$12.92 \AA.
Taking into account the SOC, the obtained LDA\cite{SQE09} band dispersions is shown in Fig.~\ref{figS1}a for the undistorted Sr$_2$IrO$_4$.

Next, we construct a two-dimensional TB+SOC model for the low-energy band structure using five localized 5$d$ Wannier orbitals centered at an Ir site labeled by $\alpha,\beta= 1 (d_{yz})$, $2 (d_{zx})$, $3 (d_{xy})$, $4 (d_{3z^2-r^2})$, $5 (d_{x^2-y^2})$,
\begin{align}
\tilde{\mathcal{H}}_0&=\sum_{ij,\alpha\beta,\sigma} \tilde{t}^{\alpha\beta}_{ij} d_{i\alpha\sigma}^\dagger d_{j\beta\sigma} +\sum_{i\alpha\sigma}\epsilon_\alpha d^\dagger_{i\alpha\sigma} d_{i\alpha\sigma} \nonumber \\
&+\lambda_{\rm soc} \sum_{i,\alpha\beta,\sigma\sigma^\prime} \left< \alpha \left|\mathbf{L}\right| \beta\right> \cdot \left< \sigma\left| \mathbf{S}\right| \sigma^\prime\right> d_{i\alpha\sigma}^\dagger d_{i\beta\sigma^\prime}.
\label{Seq:tb0}
\end{align}
Here $d_{i\alpha\sigma}^\dagger$ creates an electron with spin-$\sigma$ in the $\alpha$-th orbital at site $i$.
It is important to note that in Eq.~(\ref{Seq:tb0}), the spin and orbital are both defined in the {\em global} coordinates ($x,y,z$) of the system.
The hopping integrals $\tilde{t}_{ij}^{\alpha\beta} \equiv \tilde{t}_{\alpha\beta} [x_j-x_i, y_j-y_i]$ are {\it real} and given in Table \ref{tab:ts} for up to fifth nearest neighbors.
The obtained crystalline electric field (CEF) hierarchy is given by $\epsilon_{1,\cdots,5} =(0, 0, 202, 3054, 3831)$ meV.
The last term in Eq.~(\ref{Seq:tb0}) describes the atomic SOC and the matrix representation is given explicitly in Eq. (\ref{SOC:d}).
The LDA value for the atomic SOC is $\lambda_{\rm soc}=357$ meV.
The band dispersion produced by the TB+SOC Hamiltonian is shown in Fig.~\ref{figS1}b, which captures faithfully the low-energy part of the LDA band structure in Fig.~\ref{figS1}a.

\begin{table}
\caption{Hopping integrals $\tilde{t}_{\alpha\beta}[\Delta x, \Delta y]$
in units of meV.
$[\Delta x, \Delta y]$ denotes the in-plain hopping vector,
and $(\mu,\nu)$ the orbitals.
\textbf{I}, $\sigma_y$, and $\sigma_d$ correspond to $\tilde{t}_{\alpha\beta}[-\Delta x, -\Delta y]$, $\tilde{t}_{\alpha\beta}[\Delta x, -\Delta y]$, and $\tilde{t}_{\alpha\beta}[\Delta y, \Delta x]$, respectively, where "$\pm$" and "$\pm (\alpha',\beta')$" in the row of $(\alpha,\beta)$ mean that the corresponding hopping is equal to $\pm \tilde{t}_{\alpha\beta} [\Delta x, \Delta y]$ and $\pm \tilde{t}_{\alpha'\beta'}[\Delta x, \Delta y]$, respectively.
This table, combined with the relation $\tilde{t}_{\alpha\beta}[\Delta x, \Delta y] = \tilde{t}_{\beta\alpha}[-\Delta x, -\Delta y]$, gives all the $ab$-plane hoppings $\geq 1$meV up to fifth neighbors.}
\begin{tabular}{c|rrrrr|ccc}
($\alpha,\beta$) &[1,0]&[1,1]&[2,0]&[2,1]&[2,2]& \textbf{I}& $\sigma_y$ &$\sigma_d$\\
\hline
(1,1) &$-$66    &16     &12    &5       &2      & +   &+  &+(2,2)  \\
(1,2) &         &3      &      &1       &$-$2   & +   &$-$&+       \\
(2,2) &$-$391   &16     &30    &6       &2      & +   &+  &+(1,1)  \\
(3,3) &$-$391   &$-$139 &30    &$-$10   &$-$11  & +   &+  &+       \\
(3,4) &         &88     &      &11      &9      & +   &$-$&+       \\
(3,5) &         &       &      &$-$12   &       & +   &$-$&$-$     \\
(4,4) &$-$245   &$-$38  &$-$6  &$-$3    &$-$1   & +   &+  &+       \\
(4,5) &309      &       &30    &8       &       & +   &+  &$-$     \\
(5,5) &$-$792   &173    &$-$140&3       &       & +   &+  &+       \\
\end{tabular}
\label{tab:ts}
\end{table}

\begin{figure}
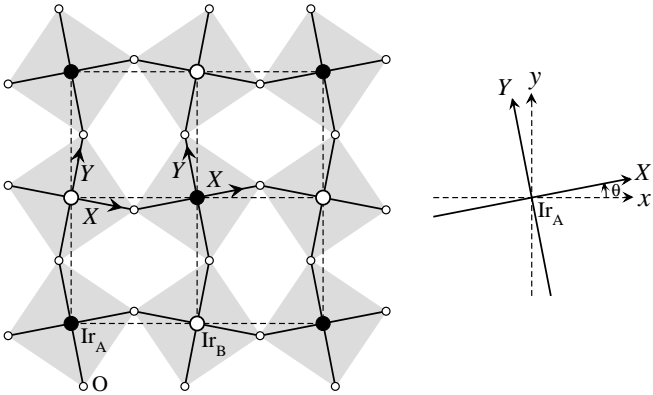

\begin{center}
\fig{3.4in}{figS2.eps}
\caption{Schematic picture of a IrO$_2$ layer.
Large filled or open circles denote Ir atoms on two sublattices, and small open circles are oxygen atoms.
Small $x$, $y$ are the global cubic axis, while capital $X$, $Y$ denote local sublattice-dependent axis.
The IrO$_6$ octahedra of sublattice A/B are rotated about $c$-axis by $\theta=\pm 11^\circ$.}
\label{figS2}
\end{center}
\end{figure}

In realistic Sr$_2$IrO$_4$ materials, the corner-shared IrO$_6$ are not aligned but rotated clockwise and anticlockwise alternately around the $c$-axis by about $11^\circ$.
The top view of the staggered octahedral rotation is shown schematically in Fig.~\ref{figS2}, which lowers the symmetry group of Sr$_2$IrO$_4$ from $I4/mmm$ to $I4_1 /acd$ and the corresponding conventional unit cell contains four ($\sqrt{2}\times\sqrt{2}\times 2$) formula units.
The lattice constants we used here are taken from the values obtained at 295 K by Crawford \textit{et. al.} \cite{Scrawford94}: $a=b=$5.497 \AA ($a_t=$3.887 \AA), $c=$25.798 \AA ($c_t=$12.899 \AA).
The obtained LDA band structure near the Fermi level is shown in Fig.~\ref{figS1}c in the reduced zone in two dimensions (2D) plotted in the inset in Fig. \ref{figS1}a since the staggered octahedral rotation doubles the 2D unit cell.
It clearly deviates from the grey lines in Fig.~\ref{figS1}d obtained by simply folding the band structure of the TB+SOC model in Eq.~(\ref{Seq:tb0}) into the 2D reduced BZ.
The main difference is that the band crossing (mainly $d_{xy}$ and $d_{x^2-y^2}$ orbitals) above the Fermi level near $\Gamma$ is lifted by the structure distortion.

In order to obtain the tight-binding model in the presence of the structure distortion, it is essential to transform the atomic spin/orbital into the {\em local} coordinates that follow the staggered octahedral rotation.
The SOC term is invariant as shown in the last section.
Assuming the atomic crystal fields in the local IrO$_6$ environment are unchanged to a good approximation, the main effects of the octahedral rotation is on the hopping term in Eq. (\ref{Seq:tb0}).
This amounts to a spatial rotation from the global ($x,y,z$) to the local ($X,Y,Z$) coordinates shown in Fig.~\ref{figS2} by angles $\theta_i$ and a spin rotation by angles $\theta_i$ \cite{Sfawang11}, i.e. ${\cal R}_i =e^{-iL_z\theta_i} \otimes e^{iS_z\theta_i}$.
The hopping of an electron from site $j$ to site $i$ is described by the $10\times 10$ matrix, $\tilde{t}_{ij}$, in the spin-orbital space.
Carrying out this rotation on the hopping matrix, $t_{ij} = {\cal R}_i^\dagger \tilde{t}_{ij} {\cal R}_j$, one finds that the hopping matrix $t_{ij}$ is in general {\it complex} and spin-orbital dependent but diagonal in spin space.
The TB+SOC model in the local ($X,Y,Z$) coordinates is thus given by
\begin{align}
\mathcal{H}_0&=\sum_{ij,\mu\nu,\sigma} t^{\mu\nu,\sigma}_{ij} d_{i\mu\sigma}^\dagger d_{j\nu\sigma}+ \sum_{i\mu\sigma}\epsilon_\mu
d^\dagger_{i\mu\sigma}d_{i\mu\sigma} \nonumber \\
&+\lambda_{\rm soc} \sum_{i,\mu\nu,\sigma\sigma^\prime} \left< \mu\left|\mathbf{L}\right|\nu\right> \cdot \left< \sigma\left| \mathbf{S}\right| \sigma^\prime\right> d_{i\mu\sigma}^\dagger d_{i\nu\sigma^\prime},
\label{Seq:tb}
\end{align}
where the orbital indices $\mu,\nu= 1 (d_{YZ})$, $2 (d_{ZX})$, $3 (d_{XY})$, $4 (d_{3Z^2-R^2})$, $5 (d_{X^2-Y^2})$, as given in Eq.~(1) in the main text. The obtained band dispersion is shown as black lines in Fig.~\ref{figS1}d, which provides an accurate five-orbital TB+SOC description of the LDA band structure in the presence of lattice distortion shown in Fig.~\ref{figS1}c.

\subsection{The $J_{\rm eff}=1/2$ doublet}

Since the cubic crystal field $\Delta_c=10Dq$ ($\simeq 3.4$ eV) is much larger than $\lambda_{\rm soc}$, the five $d$ atomic orbitals split into the 3-fold $t_{2g}$ manifold consisting of $d_{YZ}$, $d_{ZX}$, and $d_{XY}$, and the higher lying 2-fold $e_g$ manifold of $d_{3Z^2-R^2}$ and $d_{X^2-Y^2}$.
Projecting out the high energy $e_{g}$ manifold, it has been shown \cite{Sbleaney70, Sjackeli09} that the $t_{2g}$ manifold has an effective angular moment $\ell=1$ with triplets $|1,0\rangle =d_{XY}$, $|1,\pm 1\rangle = -\frac{1}{\sqrt{2}}(id_{ZX}\pm d_{YZ})$ and reversed angular momentum $\textbf{L} \rightarrow -\textbf{L}$.
The truncated matrix form of the SOC in the $t_{2g}$ manifold thus reads
\begin{equation}
\mathcal{H}_\text{SOC}= -\lambda \textbf{L} \cdot \textbf{S} =
\frac{\lambda}{2} \left( \begin{array}{ccc|ccc}
0 & -i & 0 & 0 & 0 & 1 \\
i & 0 & 0 & 0 & 0 & i \\
0 & 0 & 0 & -1 & -i & 0 \\
\hline
0 & 0 & -1 & 0 & i & 0 \\
0 & 0 & i & -i & 0 & 0 \\
1 & -i & 0 & 0 & 0 & 0 \end{array}\right)
\label{SOC:t2g}
\end{equation}
in the basis of ($d_{YZ,\downarrow}$, $d_{ZX,\downarrow}$, $d_{XY,\downarrow}$; $\downarrow \rightarrow \uparrow$)$^T$.
The SOC further splits the $t_{2g}$ complex into spin-orbit multiplets with a $J_\text{eff}=3/2$ quartet and a higher lying $J_\text{eff}=1/2$ doublet.
The $J_\text{eff}=1/2$ doublet is given by
\begin{align}
&|\frac{1}{2}, \frac{1}{2} \rangle = \frac{1}{\sqrt{3}} \left( i d_{YZ,\downarrow} + d_{ZX,\downarrow} +id_{XY,\uparrow} \right),\text{ and } \nonumber \\
&|\frac{1}{2}, -\frac{1}{2} \rangle = \frac{1}{\sqrt{3}} \left( -i d_{YZ,\uparrow} + d_{ZX,\uparrow} +id_{XY,\downarrow} \right). \nonumber
\end{align}
%
%

\subsection{Symmetry and symmetry breaking interactions}

As discussed in the main text, there is a four-fold degeneracy in the quantum states at the band touching point (BTP) at $X$ (black lines in Figs.~2b and 4a). They correspond to a Kramers doublet protected by the nonunitary time-reversal symmetry (${\cal T}$) and the symmetry under the joint spin-orbital operation ${\cal R}_2^\prime\equiv R^\prime_s(\pi)\otimes C_2^\prime\otimes\tau_{AB}$ where $R^\prime_s(\pi)$ and $C^\prime_2$ rotate the spin and spatial coordinates/orbitals by $180$ degrees along the $[1\bar 1 0]$ direction, and $\tau_{AB}$ is an $A\leftrightarrow B$ sublattice translation. The degeneracy can be partially lifted by breaking either ${\cal T}$ or ${\cal R}_2^\prime$ separately, or completely lifted by breaking both simultaneously. Note that ${\cal T}$ is already broken by the ordered moment in the undoped canted AF insulator.

\begin{table}
\caption{Considered orders and symmetry properties.}
\begin{tabular}{c|ccc}
\qquad Order &\qquad $\tau$ &\qquad ${\cal R}^\prime_2$ & \qquad ${\cal T}$\\
\hline
Bond nematic& \qquad$\sigma_0$ & \qquad$\times$ &\qquad $\checkmark$ \\
Staggered flux&\qquad $i\sigma_0$ &\qquad $\checkmark$ &\qquad $\times$ \\
Spin nematic &\qquad $\sigma_z$ &\qquad $\checkmark$ & \qquad $\times$  \\
d-PSCO &\qquad $i\sigma_z$ &\qquad $\times$ & \qquad $\checkmark$
\end{tabular}
\label{tab:ord}
\end{table}

To determine the possible forms of the quadratic symmetry-breaking interaction, whose expectation value gives the corresponding electronic order, is a serious challenge in such multiorbital and spin-orbit coupled systems. The problem is simplified considerably by observing (1) the low-energy band dispersion is dominated by the contribution from the $J_\text{eff} =1/2$ quasiparticles (QP) defined in the previous section, and (2) the degeneracy splitting observed by ARPES is anisotropic in momentum space and consistent with having a $d$-wave form factor in both the undoped canted AF insulator and the electron-doped PM metal. The general form of the interactions can thus be expressed as
\begin{equation}
{\cal H}_\Delta = \Delta \sum_{i\in A} \sum_{j=i+\delta} S_{ij} \psi^\dagger_i \tau \psi_j + h.c.
\end{equation}
between nearest neighbors on the square lattice with $\delta=\pm \hat{x}, \pm \hat{y}$. Here the spinor $\psi_i=(\gamma_{i\uparrow}, \gamma_{i\downarrow} )^\text{T}$ with $ \gamma_{\sigma}= {1\over\sqrt{3}}\left( i\sigma d_{YZ,\bar\sigma} +d_{ZX,\bar\sigma} +id_{XY,\sigma}\right)$ the destruction operator of the $\jeff$ doublet. $S_{ij} = (-1)^{i_y+j_y}$ is the standard nearest neighbor $d$-wave form factor, and $\tau$ is a $2\times 2$ matrix in the pseudo-spin basis, which can be expressed in terms of the identity and Pauli matrices.
For simplicity, we discuss explicitly the interactions that do not flip the pseudo-spins. Pseudospin flipping interactions can be treated in a similar manner.
Therefore, there are four possible choices for $\tau$, i.e., $\tau = \sigma_0$, $i\sigma_0$, $\sigma_z$, and $i\sigma_z$.
They correspond to, respectively, $d$-wave bond nematic order \cite{jiang16}, $d$-wave circulating current or staggered flux, $d$-wave spin-nematic order, and $d$-wave pseudospin current order (d-PSCO).
They break either ${\cal T}$ or ${\cal R}^\prime_2$ symmetry as summarized in Table \ref{tab:ord}.

\begin{figure*}
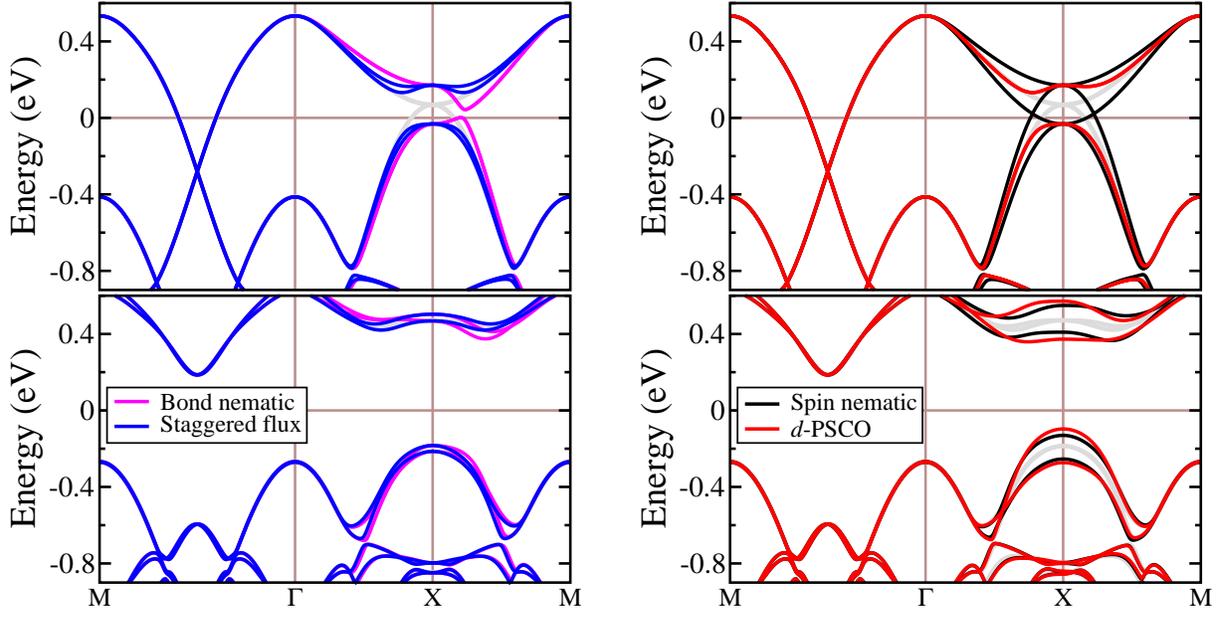

\begin{center}
\fig{3.in}{figS3a.eps} \qquad \fig{3in}{figS3b.eps}
\caption{The calculated band dispersions in the presence of, individually, bond nematic order and staggered flux (left panels), and spin nematic and d-PSCO (right panels) in the $x=0.1$ electron-doped PM metal at $(U,J)=(1.4,0)$eV (upper panels) and the undoped canted AF insulator at $(U,J)=(1.2,0.05)$eV (lower panels). The grey lines are the corresponding bands without the symmetry-breaking order. The strength of the symmetry-breaking term is $\Delta=30$meV in all cases.}
\label{figS3}
\end{center}
\end{figure*}

Fig.~\ref{figS3} shows the calculated low-energy QP band dispersions when each of the four interactions is switched on
individually with $\Delta=30$meV in both the electron doped PM metal at $x=0.1$ (top panels) and the canted AF insulator at $x=0$ (lower panels). Of the two ${\cal T}$ breaking interactions, the staggered flux can gaps out the BTP at $X$, but cannot produce the band splitting in the AF phase, whereas the $d$-wave spin nematic order splits the band degeneracy at $X$ in the canted AF phase, but fails to produce a gap for removing the hole Fermi surface pocket around $X$ in the PM metal. Note that the small band degeneracy lifting produced by the staggered flux in the canted AF phase is entirely due to the structural distortion or equivalently the canting of the AF order. Similarly, the bond nematic order does not induce the band splitting in the AF phase. Moreover it causes the QP dispersion to be strongly nematic around $X$ in the PM metal, inconsistent with experiments. It is clear from Fig.~\ref{figS3} that only the d-PSCO gives rise to an electronic structure with both the band splitting in the canted AF state and the gapping of the BTP at $X$ in the PM metal, in agreement with the observations of the ARPES experiments.


\begin{thebibliography}{99}
\bibitem{kim08}
B.J. Kim \etal, Novel ${J}_{\mathrm{eff}}=1/2$ Mott State Induced by Relativistic Spin-Orbit Coupling in Sr$_2$IrO$_4$, Phys. Rev. Lett. {\bf 101}, 076402 (2008).

\bibitem{kim09}
B.J. Kim \etal, Phase-Sensitive Observation of a Spin-Orbital Mott State in Sr$_2$IrO$_4$, Science {\bf 323}, 1329 (2009).

\bibitem{huang94}
Q. Huang \etal, Neutron Powder Diffraction Study of the Crystal Structures of Sr$_2$RuO$_4$ and Sr$_2$IrO$_4$ at Room Temperature and at 10 K, J. Solid State Chem. {\bf 112}, 355 (1994).

\bibitem{cava94}
R.J. Cava \etal, Localized-to-itinerant electron transition in Sr$_2$Ir$_{1-x}$Ru$_x$O$_4$, Phys. Rev. B {\bf 49}, 11890 (1994).

\bibitem{cao98}
G. Cao, J. Bolivar, S. McCall, J.E. Crow, and R.P. Guertin, Weak ferromagnetism, metal-to-nonmetal transition, and negative differential resistivity in single-crystal Sr$_2$IrO$_4$, Phys. Rev. B {\bf 57}, 11039(R) (1998).

\bibitem{crawford94}
M.K. Crawford \etal, Structural and magnetic studies of Sr$_2$IrO$_4$, Phys. Rev. B {\bf 49}, 9198 (1994).

\bibitem{jackeli09}
G. Jackeli and G. Khaliullin, Mott Insulators in the Strong Spin-Orbit Coupling Limit: From Heisenberg to a Quantum Compass and Kitaev Models, Phys. Rev. Lett. {\bf 102}, 017205 (2009).

\bibitem{kim12}
J.W. Kim \etal, imensionality Driven Spin-Flop Transition in Layered Iridates, Phys. Rev. Lett. {\bf 109}, 037204 (2012).

\bibitem{fujiyama12}
S. Fujiyama \etal, Two-Dimensional Heisenberg Behavior of $J_{\rm eff}=1/2$ Isospins in the Paramagnetic State of the Spin-Orbital Mott Insulator Sr$_2$IrO$_4$, Phys. Rev. Lett. {\bf 108}, 247212 (2012).

\bibitem{perkins14}
N.B. Perkins, Y. Sizyuk, and P. W\"{o}lfle, Interplay of many-body and single-particle interactions in iridates and rhodates, Phys. Rev. B {\bf 89}, 035143 (2014).

\bibitem{carter13}
J.-M. Carter, V. Shankar V., and H.-Y. Kee, Theory of metal-insulator transition in the family of perovskite iridum oxides, Phys. Rev. B {\bf 88}, 035111 (2013).

\bibitem{watanabe10}
H. Watanabe, T. Shirakawa, and S. Yunoki, Microscopic Study of a Spin-Orbit-Induced Mott Insulator in Ir Oxides, Phys. Rev. Lett. {\bf 105}, 216410 (2010).

\bibitem{watanabe14}
H. Watanabe, T. Shirakawa, and S. Yunoki, Theoretical study of insulating mechanism in multiorbital Hubbard models with a large spin-orbit coupling: Slater versus Mott scenario in Sr$_2$IrO$_4$, Phys. Rev. B {\bf 89}, 165115 (2014).

\bibitem{arita12}
R. Arita, J. Kune\v{s}, A.V. Kozhevnikov, A.G. Eguiluz, and M. Imada, \textit{Ab initio} Studies on the Interplay between Spin-Orbit Interaction and Coulomb Correlation in Sr$_2$IrO$_4$ and Ba$_2$IrO$_4$, Phys. Rev. Lett. {\bf 108}, 086403 (2012).

\bibitem{hsieh12}
D. Hsieh, F. Mahmood, D.H. Torchinsky, G. Cao, and N. Gedik, Observation of a metal-to-insulator transition with both Mott-Hubbard and Slater characteristics in Sr$_2$IrO$_4$ from time-resolved photocarrier dynamics, Phys. Rev. B {\bf 86}, 035128 (2012).

\bibitem{jin09}
H. Jin, H. Jeong, T. Ozaki, and J.-J. Yu, Anisotropic exchange interactions of spin-orbit-integrated states in Sr$_2$IrO$_4$, Phys. Rev. B {\bf 80}, 075112 (2009).

\bibitem{liu15}
P.-T. Liu \etal, Anisotropic magneti couplings and structure-driven canted to collinear transitions in Sr$_2$IrO$_4$ by magnetically constrained moncollinear DFT, Phys. Rev. B {\bf 92}, 054428 (2015).

\bibitem{pliu16}
P.-T. Liu \etal, Electron and hole doping in the relativistic Mott insulator Sr$_2$IrO$_4$: A first-principles study using band unfolding technique, Phys. Rev. B {\bf 94}, 195145 (2016).

\bibitem{fawang11}
Fa Wang and T. Senthil, Twisted Hubbard Model for Sr$_2$IrO$_4$: Magnetism and Possible High Temperature Superconductivity, Phys. Rev. Lett. {\bf 106} 136402 (2011).

\bibitem{meng14}
Z.-Y. Meng, Y.B. Kim, and H.-Y. Kee, Odd-Parity Triplet Superconducting Phase in Multiorbital Materials with a Strong Spin-Orbit Coupling: Application to Doped Sr$_2$IrO$_4$, Phys. Rev. Lett. {\bf 113}, 177003 (2014).

\bibitem{torre15}
A. de la Torre \etal, Collapse of the Mott Gap and Emergence of a Nodal Liquid in Lightly Doped Sr$_2$IrO$_4$, Phys. Rev. Lett. {\bf 115}, 176402 (2015).

\bibitem{moser14}
S. Moser \etal, The electronic structure of the high-symmetry perovskite iridate Ba$_2$IrO$_4$, New J. Phys. {\bf 16}, 013008 (2014).

\bibitem{uchida14}
M. Uchida \etal, Correlated vs. conventional insulating behavior in the J$_\text{eff}$ = $\frac{1}{2}$ vs. $\frac{3}{2}$ bands in the layered iridate Ba$_2$IrO$_4$, Phys. Rev. B {\bf 90}, 075142 (2014).

\bibitem{brouet15}
V. Brouet \etal, Transfer of spectral weight across the gap of Sr$_2$IrO$_4$ induced by La doping, Phys. Rev. B {\bf 92}, 081117(R) (2015).

\bibitem{nie15}
Y.F. Nie \etal, Interplay of Spin-Orbit Interactions, Dimensionality, and Octahedral Rotations in Semimetallic Sr$_2$IrO$_4$, Phys. Rev. Lett. {\bf 114}, 016401 (2015).

\bibitem{li-cpl15}
M.-Y. Li \etal, Tuning the Electronic Structure of  Thin Films by Bulk Electronic Doping Using Molecular Beam Epitaxy, Chin. Phys. Lett. {\bf 32}, 057402 (2015).

\bibitem{korneta10}
O.B. Korneta \etal, Electron-doped Sr$_2$IrO$_{4-\delta}$ ($0\leq\delta\leq 0.04$): Evolution of a disordered $J_{\rm eff}=\frac{1}{2}$ Mott insulator into an exotic metallic state, Phys. Rev. B {\bf 82}, 115117 (2010).

\bibitem{kim-arc-14}
Y.K. Kim \etal, Fermi arcs in a doped pseudospin-1/2 Heisenberg antiferromagnet, Science {\bf 345}, 187 (2014).

\bibitem{kim-dpg-16}
Y.K. Kim, N.H. Sung, J.D. Denlinger, and B.J. Kim, Observation of a $d$-wave gap in electron-doped Sr$_2$IrO$_4$, Nat. Phys. {\bf 12}, 37 (2016).

\bibitem{dlfeng15}
Y.J. Yan \etal, Electron-Doped Sr$_2$IrO$_4$: An Analogue of Hole-Doped Cuprate Superconductors Demonstrated by Scanning Tunneling Microscopy, Phys. Rev. X {\bf 5}, 041018 (2015).

\bibitem{zhao15}
L. Zhao \etal, Evidence of an odd-parity hidden order in a spin-orbit coupled correlated iridate, Nat. Phys. {\bf 12}, 32 (2015).

\bibitem{jeong17}
J. Jeong, Y. Sidis, A. Louat, V. Brouet, and P. Bourges, Time-reversal symmetry breaking hidden order in Sr$_2$(Ir,Rh)O$_4$, Nat. Comm. {\bf 8}, 15119 (2017).

\bibitem{martins11}
C. Martins, M. Aichhorn, L. Vaugier, and S. Biermann, Reduced Effective Spin-Orbital Degeneracy and Spin-Orbital Ordering in Paramagnetic Transition-Metal Oxides: Sr$_2$IrO$_4$ versus Sr$_2$RhO$_4$, Phys. Rev. Lett. {\bf 107}, 266404 (2011).

\bibitem{QE09}
P. Giannozzi \etal, QUANTUM ESPRESSO: a modular and open-source software project for quantum simulations of materials, J. Phys.: Condens. Matter {\bf 39}, 395502 (2009).




\bibitem{angle}
We have checked that using the value of $11^\circ$ for the canting angle, which is close to the measured value, does not give any appreciable changes in the obtained band dispersion.

\bibitem{kivelson}
E. Berg, C.-C. Chen, and S.A. Kivelson, Stability of Nodal Quasiparticles in Superconductors with Coexisting Orders, Phys. Rev. Lett. {\bf 100}, 027003 (2010).

\bibitem{jiang16}
K. Jiang, J.-P. Hu, H. Ding, and Z. Wang, Interatomic Coulomb interaction and electron nematic bond order in FeSe, Phys. Rev. B {\bf 93}, 115138 (2016).

\bibitem{liu16}
X. Liu \etal, Anisotropic softening of magnetic excitations in lightly electron-doped Sr$_2$IrO$_4$, Phys. Rev. B {\bf 93}, 241102(R) (2016).

\bibitem{gretarsson16}
H. Gretarsson \etal, Persistent Paramagnons Deep in the Metallic Phase of Sr$_{2-x}$La$_x$IrO$_4$, Phys. Rev. Lett. {\bf 117}, 107001 (2016).

\bibitem{jxli15}
Hu Wang, Shun-Li Yu, and Jian-Xin Li, Fermi arcs, pseudogap, and collective excitations in doped Sr$_2$IrO$_4$: A generalized fluctuation exchange study, Phys. Rev. B {\bf91}, 165138 (2015).




\bibitem{pan01}
S.H. Pan \etal, Microscopic electronic inhomogeneity in the high-Tc superconductor Bi$_2$Sr$_2$CaCu$_2$O$_{8+x}$, Nature {\bf 413}, 282 (2001).

\bibitem{zqwang02}
Z. Wang, J.R. Engelbrecht, S.C. Wang, H. Ding, and S.H. Pan, Inhomogeneous $d$-wave superconducting state of a doped Mott insulator, Phys. Rev. B {\bf 65}, 064509 (2002).

\bibitem{mcelroy05}
K. McElroy \etal,
Atomic-Scale Sources and Mechanism of Nanoscale Electronic Disorder in Bi$_2$Sr$_2$CaCu$_2$O$_{8+\delta}$, Science {\bf 309}, 1048 (2005).

\bibitem{sen07}
Sen Zhou, Hong Ding, and Ziqiang Wang, Correlating Off-Stoichiometric Doping and Nanoscale Electronic Inhomogeneity in the High-T$_c$ Superconductor Bi$_2$Sr$_2$CaCu$_2$O$_{8+\delta}$, Phys. Rev. Lett. {\bf 98}, 076401 (2007).

\bibitem{hoffman12}
I. Zeljkovic \etal,
Imaging the Impact of Single Oxygen Atoms on Superconducting Bi$_2$Sr$_2$CaCu$_2$O$_{8+x}$, Science {\bf 337}, 320 (2012).

\bibitem{ren09}
Liang Ren Niestemski and Ziqiang Wang, Valence Bond Glass Theory of Electronic Disorder and the Pseudogap State of High-Temperature Cuprate Superconductors, Phys. Rev. Lett. {\bf 102}, 107001 (2009).

\bibitem{disorder}
We average over 40 disorder realizations on samples of 20$\times$20 sites. To reduce the finite size effects, for each disorder realization, we average over different boundary conditions corresponding to 10$\times$10 supercells.

\bibitem{werake}
L.K. Werake, and H. Zhao, Observation of second-harmonic generation induced by pure spin currents, Nat. Phys. {\bf 6}, 875 (2010).

\bibitem{qiu}
J. Li {\it et. al.}, Direct detection of pure ac spin current by X-ray pump-probe measurements, Phys. Rev. Lett. {\bf117}, 076602 (2016).

\bibitem{gcao15}
J.~C. Wang, \etal, Decoupling of the antiferromagnetic and insulating states in Tb-doped Sr$_2$IrO$_4$, Phys. Rev. B {\bf92}, 214411 (2015).

\bibitem{longzhang16}
L. Zhang, F. Wang, and D.-H. Lee, Compass impurity model of Tb substitution in Sr$_2$IrO$_4$, Phys. Rev. B {\bf94}, 161118(R) (2016).

\bibitem{ye13}
Feng Ye \etal, Magnetic and crystal structures of Sr$_2$IrO$_4$: A neutron diffraction study, Phys. Rev. B {\bf 87}, 140406(R) (2013).

\bibitem{dhital13}
Chetan Dhital \etal, Neutron scattering study of correlated phase behavior in Sr$_2$IrO$_4$, Phys. Rev. B {\bf 87}, 144405 (2013).

\bibitem{boseggia13}
S. Boseggia \etal, Robustness of Basal-Plane Antiferromagnetic Order and the $J_{\rm eff}=1/2$ State in Single-Layer Iridate Spin-Orbit Mott Insulators, Phys. Rev. Lett. {\bf 110}, 117207 (2013).

\bibitem{tetragonaldistortion}
D.H. Torchinsky \etal,
Structural distortion-induced magnetoelastic locking in Sr$_2$IrO$_4$ revealed through nonlinear optical harmonic generation, Phys. Rev. Lett. {\bf 114}, 096404 (2015).

\bibitem{kanemele}
C.L. Kane and E.J. Mele, $Z_2$ topological order and the quantum spin Hall effect, Phys. Rev. Lett. {\bf 95}, 146802 (2005).

\bibitem{norman}
S.D. Matteo and M.R. Norman, Magnetic ground state of Sr$_2$IrO$_4$ and implications for second-harmonic generation, Phys. Rev. B {\bf 94}, 075148 (2016).

\bibitem{raghu}
S. Raghu, X.-L. Qi, C. Honerkamp, and S.-C. Zhang, Topological Mott Insulators, Phys. Rev. Lett. {\bf100}, 156401 (2008).

\bibitem{kane}
S.M. Young and C.L. Kane, Dirac semimetals in two dimensions, Phys. Rev. Lett. {\bf115}, 126803 (2015).

\end{thebibliography}

\begin{thebibliography}{99}
\bibitem{wiki}
https://en.wikipedia.org/wiki/Table\_of\_spherical\_harmonics

\bibitem{Srandall57}
J.J. Randall, Lewis Katz, and Roland Ward, The Preparation of a Strontium-Iridium Oxide Sr$_2$IrO$_4$. J. Am. Chem. Soc. \textbf{79}, 266 (1957).

\bibitem{Scrawford94}
M.K. Crawford \etal, Structural and magnetic studies of Sr$_2$IrO$_4$, Phys. Rev. B {\bf 49}, 9198 (1994).

\bibitem{SQE09}
P. Giannozzi \etal, QUANTUM ESPRESSO: a modular and open-source software project for quantum simulations of materials, J. Phys.: Condens. Matter {\bf 39}, 395502 (2009).

\bibitem{Sfawang11}
Fa Wang and T. Senthil, Twisted Hubbard Model for Sr$_2$IrO$_4$: Magnetism and Possible High Temperature Superconductivity, Phys. Rev. Lett. {\bf 106} 136402 (2011).

\bibitem{Sbleaney70}
A. Abragam and B. Bleaney, Electron Paramagnetic Resonance of Transition Ions. Clarendon Press, Oxford, (1970).

\bibitem{Sjackeli09}
G. Jackeli and G. Khaliullin, Mott Insulators in the Strong Spin-Orbit Coupling Limit: From Heisenberg to a Quantum Compass and Kitaev Models, Phys. Rev. Lett. {\bf 102}, 017205 (2009).

\bibitem{jiang16}
K. Jiang, J.-P. Hu, H. Ding, and Z. Wang, Interatomic Coulomb interaction and electron nematic bond order in FeSe, Phys. Rev. B {\bf 93}, 115138 (2016).

\end{thebibliography}
\end{document}